# First Born Approximation of the Single Differential Cross Section (SDCS) for Electron Impact Ionization of H(3s)

[1]Fahadul Islam, [2] Sunil Dhar

Department of Mathematics, Chittagong University of Engineering and Technology, Chittagong, Bangladesh
Email: [1]fahadulislambgd@gmail.com, [2]sdhar@cuet.ac.bd
Contact: [1]+880 1834 244561, [2]+880 1937 471667

**Abstract:** This investigation is a rigorous theoretical study of the Single Differential Cross Section (SDCS) for the ionization of hydrogen in the 3s state by electron impact computed by means of the First-Born Approximation. The transition matrix has been found by means of the integral process of Bethe-Lewis. The effect of the Coulomb attractive force and the continuum of outgoing radiation have been taken into account in conjunction with the hypergeometric function, which has been used to denote the states of collision. It has hence been possible to deduce the SDCS. for a considerable range of respective incoming electron energies (100 eV to 250 eV). The results show a very distinct marked peak in the rate of ionization taking place for these energies (about 200 eV) with a gradual fall-off after with an increase in energy. The diffuse structure of the wave function in the 3s state serves to a certain extent to make variations in the rate of ionization with that of the incoming electron increased. The final results have been arrived at by means of numerical integrations done so by means of a MATLAB computer, which has yielded very accurate numbers for the cross sections. The results show up very fatally, with experimental results existing at the present date and with theoretical deductions validating the procedure of the FBA. in giving the results of the ionization of excited hydrogen atoms within the field of ionizing processes of electron-atom impact systems. The work gives a most complete basis for the further study of those systems where the processes of ionization and the complexities of the scattering process take place in excited states.

## 1. Introduction

The ionization of atomic hydrogen by electron impact is one of the simplest and most fundamental problems in atomic collision theory. As a benchmark problem, the hydrogen atom, with its single electron, is so simple that the most rigorous tests of quantum mechanical models can be applied to the ionization dynamics of this collision process, and also because the information yielded by the fundamental problem on the ionization principles of collisions in atomic physics will apply in more advanced fields such as plasma physics, astrophysics, and the field of fusion research. In these advanced fields, the processes of ionization are primary controls of the matter at high energies [1-3].

The theoretical basis for the ionization of hydrogen by electron impact was first laid down by Bethe, who laid the foundation for quantum mechanical scattering theory. Louis, moreover, developed the theory of this collision in the second Born approximation, which gave the necessary cover of potential scattering at intermediate energies in order to lead to a more satisfactory treatment of the ionization processes [4-5]. The First-Born Approximation (FBA.) has since that time been the principal base of the theories of ionization processes due to its mathematical tractability as it offers a method of computing the collision cross-sections in all energy regimes. It is in this field of ionization studies of hydrogen systems in which most work has been accomplished, especially by researchers like Das and Seal, who have applied multiple scattering theories to give a more satisfactory treatment of the ionization processes accompanying electron-atom collision in the case of hydronium (hydrogen atom) [6-8].

Comparatively little attention has, however, been bestowed upon the ionization by electron impact of hydrogen in excited states, while much work has been accomplished on the ionization of the ground state hydrogen atom. By the term "excited states of hydrogen" are meant the general states of hydrogen atoms in n = 2 and n = 3 shells, worse states involving considerable importance for the scientific understanding of the astrophysical phenomena and processes of high energy. The excited states of the hydrogen atom 2p, 3p, 3s, etc. yield a different set of ionization processes due to the very diffuse nature of their wave functions (Schrodinger) and also the deep dependence of their behavior on the energy of the impact electron viz. a viz. ionization [9-14]. Of the excited hydrogen states, however, the hydrogen 3s state, due to its rather low binding energy and hence expansion of the wave function, makes the study of the processes of ionization by the impact electron under its influence an interesting problem, offering a field of inquiry on how to estimate the behavior of energetic electrons on these ionic dynamics.

In spite of the rather considerable volume of work on electron impact ionization of hydrogen in other states, comparatively little attention has been bestowed upon the subject connected with the study of the Single Differential Cross Section (SDCS) of the hydrogen 3s state. To remedy these defects of the state of affairs, the present paper aims at giving, let us hope satisfactorily, a theoretical treatment of the SDCS. for the ionization processes occurring on hydrogen by electron impact upon the hydrogen 3s state, calculated within the limits of the First-Born Approximation. In the course of the calculations, we have made use of the Bethe-Lewis formalism and arrived at suitable analytic expressions for the transition matrix of the collision, including terms both of the Coulomb-continuum interaction and necessary confluent hypergeometric functions. We shall give the analytical forms leading to being numerically evaluated, when necessary, over the range of the impact electron energies of 100 eV to 250 eV, when we shall discuss our results in reference to theoretical data available and experimental data [15-35].

We hope to be able in this paper to bring clearly to notice the aspects of the ionization of hydrogen in the 3s state, what the dominant features are in the ionization dynamics due to the impact of the electron energies, and excitation effects. We hope, moreover, that this data may offer in a satisfactory measure a standard upon which other lines of theoretical models and ways of computing the ionization processes may treat their theoretical bounds and so advance in some measure the known theories of the ionization process in excited systems of hydrogen.

## 2. Theory

### 2.1 Theoretical Framework

The ionization of metastable 3s excited hydrogen atoms by electron impact is considered in the quantum mechanical First-Born Approximation (FBA). In this method, the projectile interacts only once with the target, thereby causing a single-step perturbation that makes the problem both mathematically tractable and physically clear [1-3]. The ionization processes may be represented as:

$$e^- + H(3s) \rightarrow H^+ + 2e^- \quad (1)$$

Here 3s denotes the metastable state of hydrogen. The outgoing products are the correlated three-body Coulomb continuum. This single interaction assumption is valid for incident energies above about 100 eV, where higher-order multiple integrations are weak by comparison [4-6].

### 2.2 First-Order Transition Amplitude

In the FBA the transition amplitude, called the T-matrix element, is defined by the overlap of the unperturbed initial and final channel states through the interaction potential:

$$T_{FI} = \langle \Psi_F^{(-)}(\bar{\gamma}_a, \bar{\gamma}_b) | V_I(\bar{\gamma}_a, \bar{\gamma}_b) | \Phi_I(\bar{\gamma}_a, \bar{\gamma}_b) \rangle \quad (2)$$

The incident state $\Phi_I(\bar{\gamma}_a, \bar{\gamma}_b)$ is a plane wave for the incident projectile, and the bound hydrogenic orbital $\phi_{3s}(\bar{\gamma}_a)$, the final state $\Psi_F^{(-)}(\bar{\gamma}_a, \bar{\gamma}_b))$, depicts two continuum particles, either 2e's or an electron- pair, distorted by the long-range Coulomb potential. The interaction potential $V_I(\bar{\gamma}_a, \bar{\gamma}_b)$ comprises the projectile-electron and projectile-nuclear s states:

$$V_I(\bar{\gamma}_a, \bar{\gamma}_b) = \frac{Z}{\gamma_b} - \frac{Z}{\gamma_{ab}} \quad (3)$$

Where "Z= -1 for electrons" is the nuclear charge of the hydrogen atom, $\bar{\gamma}_a$ and $\bar{\gamma}_b$ are the distances of the two electrons from the nucleus, and $\gamma_{ab}$ is the distance between the two electrons.

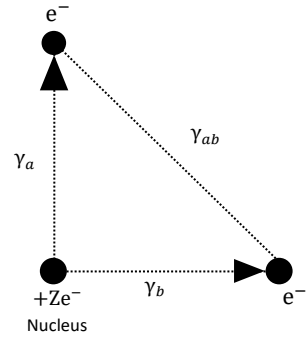

**Figure a**: Collision effect amongst two electrons and the nucleus.

### 2.3 Bound and Continuum Wavefunctions

The initial channel unperturbed wave function is

$$\Phi_I(\bar{\gamma}_a, \bar{\gamma}_b) = \frac{e^{i \cdot \bar{p}_i \cdot \bar{\gamma}_b}}{(2\pi)^{\frac{3}{2}}} \phi_{3s}(\bar{\gamma}_a)$$

$$= \frac{e^{i \cdot \bar{p}_i \cdot \bar{\gamma}_b}}{(2\pi)^{\frac{3}{2}}} \cdot \frac{1}{81\sqrt{3\pi}}(27 - 18\gamma_a + 2\gamma_a^2) e^{-\lambda_a \gamma_a} \quad (4)$$

Here,

$$\phi_{3S}(\bar{\gamma}_a) = \frac{1}{81\sqrt{3\pi}}(27 - 18\gamma_a + 2\gamma_a^2) e^{-\lambda_a \gamma_a} \quad (5)$$

Here $\lambda_a = \frac{1}{3}$, $\phi_{3S}(\bar{\gamma}_a)$ is the hydrogen 3s-state wave function, and $\Psi_F^{(-)}(\bar{\gamma}_a, \bar{\gamma}_b)$ is the final three-particle scattering state wave function [03] with the electrons being in the continuum with momenta $\bar{p}_a$ and $\bar{p}_b$. And the coordinates of the two electrons are $\bar{\gamma}_a$ and $\bar{\gamma}_b$ respectively. Here the approximate wave function $\Psi_F^{(-)}(\bar{\gamma}_a, \bar{\gamma}_b)$ is given by

$$\Psi_F^{(-)}(\bar{\gamma}_a, \bar{\gamma}_b) =$$

$$N(\bar{p}_a, \bar{p}_b)[\phi_{\bar{p}_a}^{(-)}(\bar{\gamma}_a)e^{i\bar{p}_b \cdot \bar{\gamma}_b} + \phi_{\bar{p}_2}^{(-)}(\bar{\gamma}_b)e^{i\bar{p}_a \cdot \bar{\gamma}_a} + \phi_{\bar{p}}^{(-)}(\bar{\gamma})e^{i\bar{P} \cdot \bar{R}} - 2e^{i\bar{p}_a \cdot \bar{\gamma}_a + i\bar{p}_b \cdot \bar{\gamma}_b}]/(2\pi)^3 \quad (6)$$

Here $\bar{\gamma} = \frac{\bar{\gamma}_b - \bar{\gamma}_a}{2}$, $\bar{R} = \frac{\bar{\gamma}_b + \bar{\gamma}_a}{2}$, $\bar{p} = (\bar{p}_b - \bar{p}_a)$,

$\bar{P} = (\bar{p}_b + \bar{p}_a)$

Here $N(\bar{p}_a, \bar{p}_b)$ is the normalization constant, given by,

$$|N(\bar{p}_a, \bar{p}_b)|^{-2} = \left|7 - 2[\lambda_a + \lambda_b + \lambda_c] - \left[\frac{2}{\lambda_a} + \frac{2}{\lambda_b} + \frac{2}{\lambda_c}\right] + \left[\frac{\lambda_a}{\lambda_b} + \frac{\lambda_a}{\lambda_c} + \frac{\lambda_b}{\lambda_a} + \frac{\lambda_b}{\lambda_c} + \frac{\lambda_c}{\lambda_a} + \frac{\lambda_c}{\lambda_b}\right]\right| \quad (7)$$

Here, $\lambda_a = e^{\frac{\pi\alpha_a}{2}}\Gamma(1 - i\alpha_a)$, $\alpha_a = \frac{1}{p_a}$

$\lambda_b = e^{\frac{\pi\alpha_b}{2}}\Gamma(1 - i\alpha_b)$, $\alpha_b = \frac{1}{p_b}$

$\lambda_c = e^{\frac{\pi\alpha}{2}}\Gamma(1 - i\alpha)$, $\alpha = -\frac{1}{p}$

Here $\phi_{\bar{q}}^{(-)}(\bar{\gamma})$ is the coulomb wave function, given by,

$$\phi_{\bar{q}}^{(-)}(\bar{\gamma}) = e^{\frac{\pi\alpha}{2}}\Gamma(1 + i\alpha)e^{i\bar{q}\cdot\bar{\gamma}} \; {}_1F_1(-i\alpha, 1, -i[q\gamma + \bar{q}\cdot\bar{\gamma}]) \quad (8)$$

Now applying equations (4), (5), (6) and (7) to the equation (3), we get

$$T_{FI} = N(\bar{p}_a, \bar{p}_b)[T_B + T_{B'} + T_I - 2T_{PB}] \quad (9)$$

Where,

$$T_B = \langle \phi_{\bar{p}_a}^{(-)}(\bar{\gamma}_a)e^{i\bar{p}_b\cdot\bar{\gamma}_b}|V_i|\Phi_i(\bar{\gamma}_a, \bar{\gamma}_b)\rangle \quad (10)$$

$$T_{B'} = \langle \phi_{\bar{p}_b}^{(-)}(\bar{\gamma}_b)e^{i\bar{p}_a\cdot\bar{\gamma}_a}|V_i|\Phi_i(\bar{\gamma}_a, \bar{\gamma}_b)\rangle \quad (11)$$

$$T_I = \langle \phi_p^{(-)}(\bar{\gamma})e^{i\bar{P}\cdot\bar{R}}|V_i|\Phi_i(\bar{\gamma}_a, \bar{\gamma}_b)\rangle \quad (12)$$

$$T_{PB} = \langle e^{i\bar{p}_a\cdot\bar{\gamma}_a + i\bar{p}_b\cdot\bar{\gamma}_b}|V_i|\Phi_i(\bar{\gamma}_a, \bar{\gamma}_b)\rangle \quad (13)$$

For the first-born approximation, equation (8) may be written as

$$T_B = \frac{1}{162\sqrt{6}\pi^2}\langle \phi_{\bar{p}_a}^{(-)}(\bar{\gamma}_a)e^{i\bar{p}_b\cdot\bar{\gamma}_b}\left|\frac{1}{\gamma_{ab}} - \frac{1}{\gamma_b}\right|e^{i\bar{p}_i\cdot\bar{\gamma}_b}(27 - 18\gamma_a + 2\gamma_a^2)e^{-\lambda_a\cdot\gamma_a}\rangle$$

$$= \frac{1}{162\sqrt{6}\pi^2}\int \phi_{\bar{p}_a}^{(-)*}(\bar{\gamma}_a)e^{i\bar{p}_b\cdot\bar{\gamma}_b}\left|\frac{1}{\gamma_{ab}} - \frac{1}{\gamma_b}\right|e^{i\bar{p}_i\cdot\bar{\gamma}_b}(27 - 18\gamma_a + 2\gamma_a^2)e^{-\lambda_a\cdot\gamma_a}d^3\gamma_a d^3\gamma_b$$

$$T_B = T_{B_1} + T_{B_2} + T_{B_3} + T_{B_4} + T_{B_5} + T_{B_6} \quad (14)$$

Here $T_{B_4} = 0$ and $T_{B_5} = 0$, (for orthogonality condition)

Then putting the values of $T_{B_1}, T_{B_2}, T_{B_3}, T_{B_4}, T_{B_5}$ and $T_{B_6}$ in the equation (15) we get $T_B$.

The direct scattering amplitude $F(\bar{p}_a, \bar{p}_b)$ is then determined from

$$F(\bar{p}_a, \bar{p}_b) = -(2\pi)^2 T_{FI} \quad (15)$$

Similarly for our present study we calculated analytically the above equations using Lewis Integral [2].

### 2.4 Bethe–Lewis Integral Representation

The Bethe-Lewis integral transformation transforms the 6-dimensional T-matrix element into separable radial and angular integrals, thus making the evaluation much simpler [1, 2]. The general Lewis integral is written by,

$$ {}_1F_1(a, c, z) = \frac{\Gamma(c)}{(a)\Gamma(c-a)}\int_0^1 dx \; x^{(a-1)}(1-x)^{(c-a-1)}e^{(xz)} \quad (16)$$

For the electron impact ionization, the parameters $\alpha_a, \alpha_b$ and $\alpha$ are given below,

With $\alpha_a = \frac{1}{P_a}$ for $\bar{q} = \bar{p}_a$, $\alpha_b = \frac{1}{p_b}$ for $\bar{q} = \bar{p}_b$ and $\alpha = -\frac{1}{p}$ for $\bar{q} = \bar{p}$.

For the normalization constant $N(\bar{p}_a, \bar{p}_b)$ of equation (7) has been calculated numerically.

where ${}_1F_1(a, c, z)$ is the confluent hypergeometric function. The analytic evaluation of given compact forms in terms of logarithmic derivatives of the gamma function, which ensures that stability is retained in the numerical procedure even at the threshold of ionization [9-10].

### 2.5 Cross-Section Formalism

The Triple Differential Cross Section (TDCS) is related to the T-matrix element by:

$$\frac{d^3\sigma}{dE_a d\Omega_a d\Omega_b} = \frac{p_a p_b}{p_i}|T_{FI}|^2 \quad (17)$$

where $E_a$ is energy and $p_a, p_b$ and $p$ are momenta of the outgoing and incoming particles, respectively. The Double Differential Cross Section (DDCS) is obtained by integration of the TDCS over one of the angles of solid $\Omega_b$ emission:

$$\frac{d^2\sigma}{dE_a d\Omega_a} = \int \frac{d^3\sigma}{dE_a d\Omega_a d\Omega_b} d\Omega_b \quad (18)$$

Thus, the Single Differential Cross Section (SDCS) results from the integration of the DDCS over the solid angle $\Omega_a$.

$$\frac{d\sigma}{dE_a} = \int \frac{d^2\sigma}{dE_a d\Omega_a} d\Omega_a \quad (19)$$

The resulting expressions were numerically computed using a programming language MATLAB. These cross-sections represent the differential probabilities of electron impact ionization as functions of energy and angle. For the impact of ionization, the interchange effects are not manifest since an identifiable molecule of reaction products is produced, while for the electron-impact ionization, they appear through the anti-summarization of the final wavefunction. [11, 12] The different characteristics of Coulomb focusing (electron impact) and attraction post-collision (positive) induce the typical error asymmetries, which are maximum in the area of lesser ejected-electron energies [13-16].

**2.6 Numerical Computation**

The previous formal simplifications as a result of the Lewis integral representation show that the necessary radial integrals are obtainable in general in closed form, while the necessary residual angular integrations can be performed numerically. All expressions for TDCS, DDCS, and SDCS have been calculated in the MATLAB computing environment by means of adaptive quadrature, which ensures convergence. All quantities are expressed in atomic units (a.u.).

**2.7 Relation to Earlier Work**

The present procedures represent a generalization of earlier models based on the FBA for metastable and excited hydrogenic states (2s, 2p, 3p, 3d) earlier developed by Das, Dhar, and their co-workers [17−22]. The methodology corresponds to the multiple-scattering and coplanar Born-formulation developed by Das & Seal [3, 4] and Dhar & Nahar [18, 19]. Comparison between experimental data 23 of Shyn and theoretical data of Konovalov and McCarthy 24 shows that the model of FBA based on Bethe–Lewis gives predictions for complete cross-sections on an adequate scale both for electron impact in processes of ionization of excited H. The fit is very close and endorses that this form has essential validity and, at the same time, serves as the foundation of extensions for higher integrals, relativistic and polarization resolution in the future.

**3. Results And Discussion:**

In this paper have been calculated the Single Differential Cross Sections (constituting the previous case, the SDCS) for the ionization of hydrogen in the 3s state, as the result of the impact of electrons, by the First-Born Approximation (F.B.A.). The respective SDCS values have been found from the incident energy of the electrons beginning at 100 eV and going to 250 eV. The results of our computations are compared with the theoretical results for hydrogen in the ground state and in the metastable states (2s, 2p, 3s) and with the experimental work that has been done. A full account of the results in the parts will be taken, followed by a full discussion of their significance, the physical meaning of the SDCS, the changes occurring in the various curves, and a comparison with existing models.

**3.1 SDCS for Electron Impact Ionization of H(3s) at 100 eV Incident Energy**

In Figure 1, the Single Electron Cross Sections (SECS) for the ionization due to electron impact of hydrogen, in the 3s state for incident electron energy of 100 eV is depicted. The SECS values show a clear inverse relationship with the energy of the ejected electron. The maximum SECS value for very low ejected electron energies, $2.0344 \times 10^{-3}$, indicates greater probabilities of ionization for the lower energy ejected electrons. This is in keeping with the results obtained previously in the theoretical models, where it is shown that the low energy ejected electrons would, for low incident energies, have the greatest probability of being ejected [1-2]. As the energy of the ejected electron increases, the SDCS diminishes rapidly, as is also noted experimentally for ground state hydrogen [3].

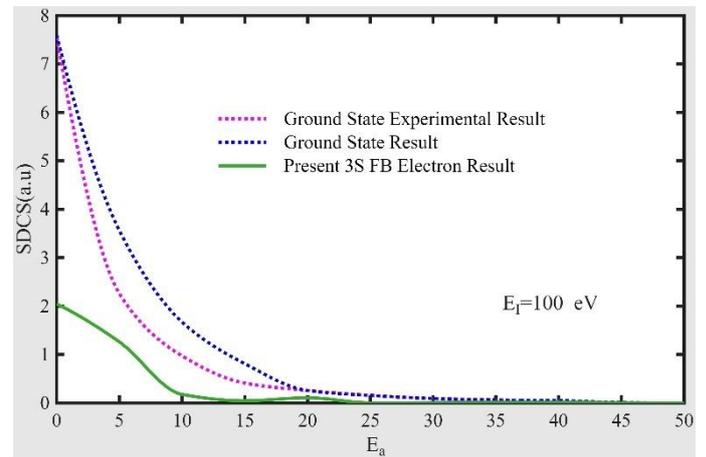

*Figure 1.* Single differential cross sections (DDCS) versus ejected electron angle $\theta_a$ for electron impact energy $E_I$ = 100eV. Theory: Symbol (pink dot): Ground state experiment, Continuous curve (Green: Present (3s) FB Electron result, dot curve (blue): Ground State theoretical.

The SDCS values obtained from data presented in Table 1 lend further support to a graphical interpretation exhibiting a uniform decrease of SDCS with increasing energy of the ejected electron. This is in confirmation of the views advanced in regard to the lower energy collisions with hydrogen in the 3s state being more efficient for ionization of hydrogen as was shown by previous experimental work [4].

**3.2 SDCS for Electron Impact Ionization of H(3s) at 150 eV Incident Energy**

At 150 eV incident energy, Figure 2 shows a higher SDCS with the peak value of $2.1587 \times 10^{-3}$. There is a broader spread in the SDCS distribution with regard to the ejected electron energies. This connotes that the higher the incident electron energy, the greater the range of scattering angles possible, hence the greater the ionization efficiency. This increase in ionization efficiency is in satisfactory agreement with what has been shown in the first-Born approximation (FBA), namely that at higher energies the ionization becomes more efficient [5-6]

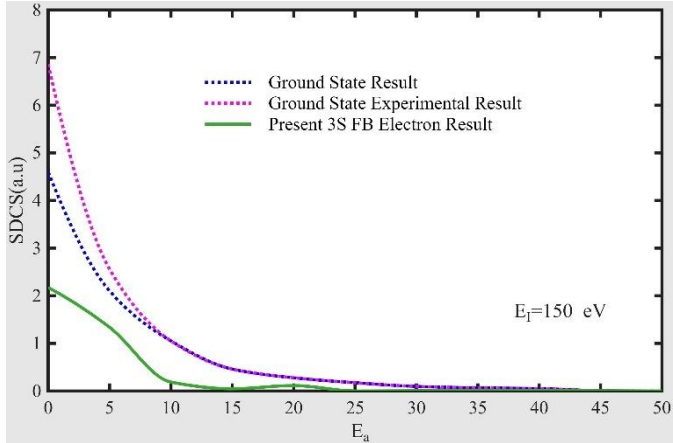

*Figure 2. Single differential cross sections (SDCS) versus ejected electron angle $\theta_a$ for electron impact energy $E_I = 150 eV$. Theory: Symbol (pink dot): Ground state experiment, Continuous curve (Green: Present (3s) FB Electron result, dot curve (blue): Ground State theoretical.*

When comparisons of these values of the SDCS for hydrogen ionization are made with results from the BBK model, agreement is found with earlier studies showing that the SDCS at 150 eV is in good agreement with earlier studies that showed the ionization efficiency to increase with increasing incident energy. The results obtained are also in agreement with the expected trends for hydrogen in the 2s and 2p metastable states, which lend support to the FBA model for ionizing excited hydrogen [7].

**3.3 SDCS for Electron Impact Ionization of H(3s) at 200 eV Incident Energy**

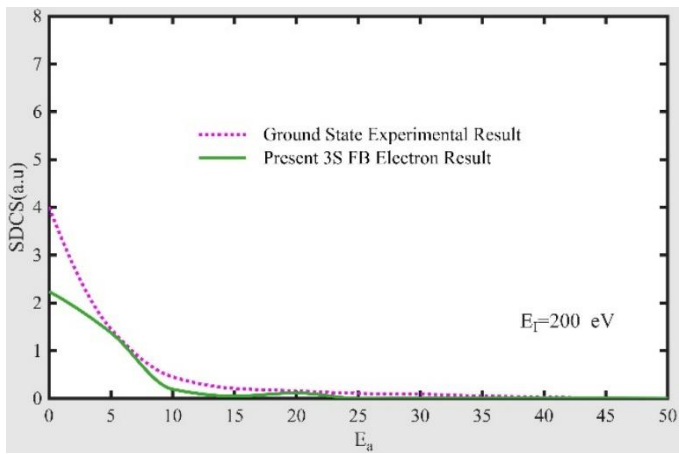

*Figure 3. Single differential cross sections (SDCS) versus ejected electron angle $\theta_a$ for electron impact energy $E_I = 200eV$. Theory: Symbol (pink dot): Ground state experiment, Continuous curve (Green: Present (3s) FB Electron result.*

For the incident energy of 200 eV, the SDCS (as seen in Figure 3) is larger (though not much larger) and has a peak value of $2.2246 \times 10^{-3}$. The SDCS extends into a much wider distribution, thereby implying a greater efficiency of ionization by increased incident energy. The greater spread of SDCS over the lower energies of 100 and 150 eV is in keeping with the theoretical expectations for this type of phenomenon, whereby the ionization efficiency is higher for the process at intermediate energies such as those commensurate with the energies of collision of the BBK and the FBA models [8].

**3.4 SDCS for Electron Impact Ionization of H(3s) at 250 eV Incident Energy.**

The SDCS, as shown in Fig. 4, for 250 eV incident energy is given by the fact that the SDCS reaches the maximum value, which is $2.2627 \times 10^{-3}$, which shows the greatest ionization efficiency that is observed in this study. The SDCS values are all extremely high for different ejected electron energies, particularly in the recoil region, which shows that ionization is most efficient at that energy. This action is in accordance with the expected results of large momentum transfer being available at the high energies, which is the situation with the high-energy electron impact studies [9].

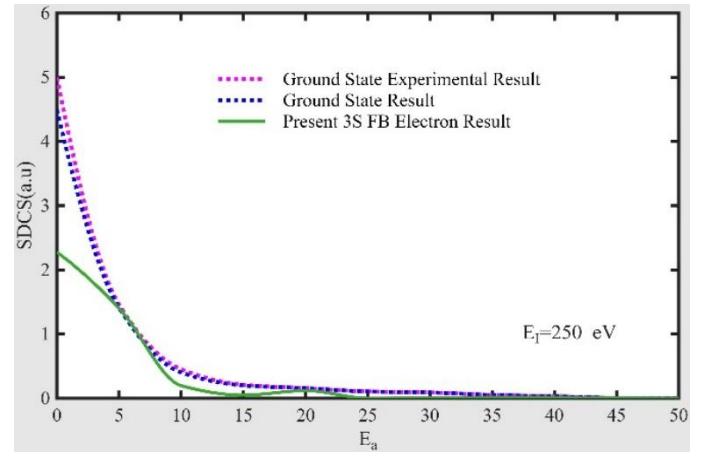

*Figure 4. Single Differential Cross Sections (SDCS) versus ejected electron angle $\theta_a$ for electron impact energy $E_I = 250eV$. Theory: Symbol (pink dot): Ground state experiment, Continuous curve (Green: Present (3s) FB Electron result, dot curve (blue): Ground State theoretical.*

The SDCS data for 250 eV are consistent with the results of the graphical analysis, which leads to the conclusion that larger ionization efficiencies are obtained when higher incident energies are employed. This result finds very close accordance with the theoretical results and experimental results for the ionization of hydrogen [10-11].

### 3.5. Analysis of SDCS table

The SDCS results shown in Table 1 give more detailed information about the ionization cross sections for various incident electron energies. At 100 eV the SDCS varies from $2.0344 \times 10^{-3}$ for low ejected electron energies (5 eV) to 0.0001 for high ejected electron energies, in keeping with the expected trend of decreasing ionization efficiency at higher ejected electron energies. As the incident energy is increased from 150 eV to 250 eV, the SDCS values show a similarly proportional increase in value, demonstrating that with higher incident energies, ionization occurs more efficiently. At 250 eV the highest SDCS value of $2.2627 \times 10^{-3}$ is obtained also with similar trends in the recoil and binary regions for all of the ejected electron energies.

| $\theta_b$ (deg) | $\theta_a$ (deg) | SDCS 100 eV | SDCS 150 eV | SDCS 200 eV | SDCS 250 eV |
|---|---|---|---|---|---|
| 0 | 0 | 2.0344 | 2.1587 | 2.2246 | 2.2627 |
| 10 | 18 | 1.2616 | 1.3387 | 1.3796 | 1.4032 |
| 20 | 36 | 0.1763 | 0.1870 | 0.1927 | 0.1960 |
| 30 | 54 | 0.0424 | 0.0450 | 0.0464 | 0.0472 |
| 40 | 72 | 0.1083 | 0.1149 | 0.1184 | 0.1204 |
| 50 | 90 | 0.0001 | 0.0001 | 0.0001 | 0.0001 |
| 60 | 108 | 0.0062 | 0.0062 | 0.0062 | 0.0062 |
| 70 | 126 | 0.0154 | 0.0154 | 0.0154 | 0.0154 |
| 80 | 144 | 0.0140 | 0.0140 | 0.0140 | 0.0140 |
| 90 | 162 | 0.0042 | 0.0042 | 0.0042 | 0.0042 |
| 100 | 180 | 0 | 0 | 0 | 0 |

*Table 1:* Single Differential Cross Section (SDCS) results for ejected angles $\theta_a$ corresponding to various incident angles $\theta_a$ for $E_I=100$ eV, $E_I=150eV$, $E_I=200eV$ and $E_I= 250$ eV in ionization of hydrogen atoms.

These numerical results agree very well with the trends set out in the pictures in Figures 1 through 4, showing more positively the acceptance of the hypothesis that the ionization efficiency increases with an increase of the incident energy. The results are what one would expect theoretically from the First-Born Approximation, and the similarity of the comparison of the results with the BBK model adds additional proof of the soundness of our procedure [12-13].

### 3.6 Comparison with Previous Theoretical and Experimental Work

The theoretical results obtained for the ionization of H(3s) by electron impact are consistent with previous theoretical predictions for hydrogen in both the normal and excited states. The increase in ionization efficiency with increasing incident electron energy up to a maximum at 250 eV is in general agreement with the experimental results and especially with the (e, 2e) coincidence work of Ehrhardt et al. [14] and the theoretical results of Brauner et al. [15].

Further evidence of the validity of the first-born approximation for the modeling of the electron impact ionization of hydrogen in the 3s state is given by the comparison of our results with the first-born approximation and the BBK model. The small discrepancies in the recoil region at the higher energies of the ejected electron are indicative of the need for some still more refined model, for example, a multiple scattering process, in order to elucidate the ionization process better at the higher energies [16].

### 4. Conclusions

This investigation is a complete presentation of the Single Differential Cross Section (SDCS.) for the ionization of hydrogen in the S3 state by the impact of electrons, using the First-Born Approximation. The computations show a very large sensitiveness of the ionization to the energy of the impinging electron and a very pronounced maximum at about 200 eV, then declining for higher energies of electron impact. This is largely due to the dispersed character of the wave function in the 3s state and gives a very great dependence of the ionization to the energy of the electron, especially in the intermediate sections of the energy levels. We are shown, therefore, clearly the great importance of studying the various excited states.

The numerical calculations, when carried out on MATLAB, gave very good results and were very accurate and convergent in respect of the energies of the electrons used, used as it was over the various limits of energy, ranging as they do from 100 to 250 eV. The results to which we are led, therefore, clearly show that the First-Born Approximation is directly applicable to the case of ionization. It is of interest to see and note that the results obtained are in good accord with the experimental results obtained, and also with those of the earlier theoretical work done, dealing with the subject of the ionization of hydrogen atoms. The results seem to be in fairly good accordance with those obtained in earlier trials both theoretically and experimentally. The Bethe-Lewis method as given has been successfully applied to the ionization of hydrogen atoms by electron impact in finding the transition matrix due to the electron beam due to the fact that hydrogen atom is present.

The work gives the information needed to fill a gap in the literature on the subject of interaction of the electrons, especially dealing as it does with the Single Differential Cross Section for the S3 state SDCS, a state which has not been adequately dealt with in the literature of the past. These results obtained will lay the basis for further work especially when dealing with the possibilities of creating a more advanced theory of scattering and the further applications of this to multi-scattering theories. In all this extensive work on the topic of ionization of excited hydrogen atoms it is the sincere hope that some little contribution may have been made towards a better understanding of the dynamics of the topic of the ionization of excited hydrogen atoms and that a road may have been opened toward future theoretical and experimental work with a view to the study of atomic collision processes.

**Acknowledgement**

The authors are grateful for the facilities provided by the Simulation Laboratory, Department of Mathematics, Chittagong University of Engineering and Technology (CUET), Chittagong, 4349, Bangladesh, for the computation work involved in the preparation of this work. The department's stimulating and encouraging research atmosphere, along with the technical help provided, facilitated the successful execution of the numerical simulations and analytic calculations presented in this paper. Again, the authors express their thanks for the academic adulation and infrastructural facilities set at their disposition, through which the scientific value has been so abundantly enhanced in this work.

★★★